\documentclass[11pt]{article}

\usepackage{graphicx}
\usepackage{color}

\makeatletter
 
\def\paragraph{\@startsection{paragraph}{4}{\z@}{+2.00ex plus
 +1ex minus +.2ex}{1.5ex plus .2ex}{\it\normalsize}}
 
\def\section{\@startsection {section}{1}{\z@}{+3.0ex plus +1ex minus
  +.2ex}{2.3ex plus .2ex}{\normalsize\bf}}
\def\subsection{\@startsection{subsection}{2}{\z@}{+2.5ex plus +1ex
minus +.2ex}{1.5ex plus .2ex}{\normalsize\bf}}
\def\subsubsection{\@startsection{subsubsection}{3}{\z@}{+3.25ex plus
 +1ex minus +.2ex}{1.5ex plus .2ex}{\normalsize\bf}}

\@addtoreset{equation}{section}

\def\appendix{\par
 \setcounter{section}{0} 
 \setcounter{subsection}{0}
 \setcounter{equation}{0}
 \def\thesection{\Alph{section}}}
 
\expandafter\ifx\csname mathrm\endcsname\relax\def\mathrm#1{{\rm #1}}\fi
 
\makeatletter

\newcount\@tempcntc
\def\@citex[#1]#2{\if@filesw\immediate\write\@auxout{\string\citation{#2}}\fi
  \@tempcnta\z@\@tempcntb\m@ne\def\@citea{}\@cite{\@for\@citeb:=#2\do
    {\@ifundefined
       {b@\@citeb}{\@citeo\@tempcntb\m@ne\@citea
        \def\@citea{,\penalty\@m\ }{\bf ?}\@warning
       {Citation `\@citeb' on page \thepage \space undefined}}%
    {\setbox\z@\hbox{\global\@tempcntc0\csname
b@\@citeb\endcsname\relax}%
     \ifnum\@tempcntc=\z@ \@citeo\@tempcntb\m@ne
       \@citea\def\@citea{,\penalty\@m}
       \hbox{\csname b@\@citeb\endcsname}%
     \else
      \advance\@tempcntb\@ne
      \ifnum\@tempcntb=\@tempcntc
      \else\advance\@tempcntb\m@ne\@citeo
      \@tempcnta\@tempcntc\@tempcntb\@tempcntc\fi\fi}}\@citeo}{#1}}

\def\@citeo{\ifnum\@tempcnta>\@tempcntb\else\@citea
  \def\@citea{,\penalty\@m}%
  \ifnum\@tempcnta=\@tempcntb\the\@tempcnta\else
   {\advance\@tempcnta\@ne\ifnum\@tempcnta=\@tempcntb \else
\def\@citea{--}\fi
    \advance\@tempcnta\m@ne\the\@tempcnta\@citea\the\@tempcntb}\fi\fi}

\newcommand{\lsim}
{\mathrel{\raisebox{-.3em}{$\stackrel{\displaystyle <}{\sim}$}}}

\def\asymp#1%
{\mathrel{\raisebox{-.4em}{$\widetilde{\scriptstyle #1}$}}}

\def\Nequal#1%
{\mathrel{\raisebox{-.5em}{$\stackrel{=}{\scriptstyle\rm#1}$}}}
\def\dsl{\mathpalette\make@slash}
\def\make@slash#1#2{\setbox\z@\hbox{$#1#2$}%
  \hbox to 0pt{\hss$#1/$\hss\kern-\wd0}\box0}

\def\beq#1\eeq{\begin{equation}#1\end{equation}}
\def\beqar{\begin{eqnarray}}
\def\eeqar{\end{eqnarray}}
\def\barr#1{\begin{array}{#1}}
\def\earr{\end{array}}
\def\bfi{\begin{figure}}
\def\efi{\end{figure}}
\def\btab{\begin{table}}
\def\etab{\end{table}}
\def\bce{\begin{center}}
\def\ece{\end{center}}
\def\nn{\nonumber}

\def\Ga{\Gamma}
\def\ga{\gamma}

\def\De{\Delta}
\def\eps{\epsilon}

\def\refeq#1{\mbox{(\ref{#1})}}

\def\reffi#1{\mbox{Fig.~\ref{#1}}}

\def\refta#1{\mbox{Table~\ref{#1}}}

\def\refse#1{\mbox{Section~\ref{#1}}}

\def\citere#1{\mbox{Ref.~\cite{#1}}}
\def\citeres#1{\mbox{Refs.~\cite{#1}}}
 
\newcommand{\TeV}{\unskip\,\mathrm{TeV}}
\newcommand{\GeV}{\unskip\,\mathrm{GeV}}

\newcommand{\pb}{\unskip\,\mathrm{pb}}

\newcommand{\ri}{{\mathrm{i}}}

\newcommand{\rd}{{\mathrm{d}}}

\newcommand{\rT}{{\mathrm{T}}}

\newcommand{\M}{{\cal{M}}}

\def\mathswitchr#1{\relax\ifmmode{\mathrm{#1}}\else$\mathrm{#1}$\fi}

\newcommand{\Pq}{\mathswitch q}
\newcommand{\PW}{\mathswitchr W}

\newcommand{\PZ}{\mathswitchr Z}

\newcommand{\Pg}{\mathswitchr g}

\newcommand{\Pp}{\mathswitchr p}

\newcommand{\Pj}{\mathswitchr j}

\newcommand{\Pep}{\mathswitchr {e^+}}
\newcommand{\Pem}{\mathswitchr {e^-}}

\newcommand{\jet}{\mathrm{jet}}
\newcommand{\Pl}{\ell}
\newcommand{\GW}{\mathswitch {\Gamma_\PW}}
\newcommand{\GZ}{\mathswitch {\Gamma_\PZ}}
 
\def\mathswitch#1{\relax\ifmmode#1\else$#1$\fi}

\newcommand{\MW}{\mathswitch {M_\PW}}

\newcommand{\MZ}{\mathswitch {M_\PZ}}

\newcommand{\GF}{\mathswitch {G_\mu}}

\newcommand{\OS}{{\mathrm{OS}}}

\def\Re{\mathop{\mathrm{Re}}\nolimits}

\newcommand{\MSbar}{{\overline{\mathrm{MS}}}}

\newcommand{\cut}{\mathrm{cut}}

\newcommand{\z}{\setbox0\hbox{+}\hbox to \wd0{\hss0\hss}}
\def\limfunc#1{\mathop{\rm #1}}

\def\Re{\limfunc{Re}}
\def\Im{\limfunc{Im}}

\def\slash#1{{\setbox0=\hbox{$#1$}
  \rlap{\ifdim\wd0>.7em\kern.22\wd0\else\kern.1\wd0\fi /}#1}}

\def\braket#1#2{\left\langle #1\vphantom{#2}
  \right. \kern-2.5pt\left| #2\vphantom{#1}\right\rangle }

\def\M{{\cal M}}

\def\cut{\mathswitchr{cut}}

\def\rT{{\mathrm{T}}}

\def\alphas{\alpha_{\mathrm{s}}}

\def\draftdate{\relax}
\def\mda{\relax}
\def\mua{\relax}
\def\mla{\relax}
\def\Mda{\relax}
\def\Mua{\relax}
\def\Mla{\relax}
\def\draft{
\def\thtystars{******************************}
\def\sixtystars{\thtystars\thtystars}
\typeout{}
\typeout{\sixtystars**}
\typeout{* Draft mode!
         For final version remove \protect\draft\space in source file *}
\typeout{\sixtystars**}
\typeout{}
\def\draftdate{\today}
\def\mua{\marginpar[\boldmath\hfil$\uparrow$]%
                   {\boldmath$\uparrow$\hfil}%
                    \typeout{marginpar: $\uparrow$}\ignorespaces}
\def\mda{\marginpar[\boldmath\hfil$\downarrow$]%
                   {\boldmath$\downarrow$\hfil}%
                    \typeout{marginpar: $\downarrow$}\ignorespaces}
\def\mla{\marginpar[\boldmath\hfil$\rightarrow$]%
                   {\boldmath$\leftarrow $\hfil}%
                    \typeout{marginpar: $\leftrightarrow$}\ignorespaces}
\def\Mua{\marginpar[\boldmath\hfil$\Uparrow$]%
                   {\boldmath$\Uparrow$\hfil}%
                    \typeout{marginpar: $\uparrow$}\ignorespaces}
\def\Mda{\marginpar[\boldmath\hfil$\Downarrow$]%
                   {\boldmath$\Downarrow$\hfil}%
                    \typeout{marginpar: $\downarrow$}\ignorespaces}
\def\Mla{\marginpar[\boldmath\hfil$\Rightarrow$]%
                   {\boldmath$\Leftarrow $\hfil}%
                    \typeout{marginpar: $\leftrightarrow$}\ignorespaces}
\overfullrule 5pt
\oddsidemargin -15mm
\marginparwidth 29mm
}

\begin{document}

\thispagestyle{empty}
\def\thefootnote{\fnsymbol{footnote}}
\setcounter{footnote}{1}
\null
\strut\hfill Cavendish-HEP-19/20, FR-PHENO-2019-009, TIF-UNIMI-2019-10
\vskip 0cm
\vfill
\begin{center}
{\Large \bf 
\boldmath{Low-virtuality photon transitions $\gamma^*\to f\bar f$ \\[.5em]
and the photon-to-jet conversion function}
\par} \vskip 2.5em
{\large
{\sc Ansgar Denner$^1$, Stefan Dittmaier$^2$, Mathieu Pellen$^3$
\\ and Christopher Schwan$^4$}\\[1.5em]
{\normalsize 
\it 
$^1$ Universit\"at W\"urzburg, 
Institut f\"ur Theoretische Physik und Astrophysik, \\ %
Emil-Hilb-Weg 22,  %
97074 W\"urzburg, %
Germany \\[.5em]
$^2$ Albert-Ludwigs-Universit\"at Freiburg, 
Physikalisches Institut, \\
Hermann-Herder-Stra\ss{}e 3,
D-79104 Freiburg, Germany \\[.5em]
$^3$University of Cambridge, Cavendish Laboratory,
        Cambridge CB3 0HE, United Kingdom \\[.5em]
$^4$ Tif Lab, Dipartimento di Fisica, Universit\`a di Milano and INFN,
Sezione di Milano,
Via Celoria 16,
20133 Milano,
Italy
}
}

\par \vskip 2em
\end{center} \par
\vskip 2cm {\bf Abstract:} \par The calculation of electroweak
corrections to processes with jets in the final state involves
contributions of low-virtuality photons leading to jets in the final
state via the singular splitting $\gamma^* \to \Pq\bar\Pq$.  These
singularities can be absorbed into a photon-to-jet ``fragmentation
function'', better called ``conversion function'', since the physical
final state is any hadronic activity rather than an identified hadron.
Using unitarity and a dispersion relation, we relate this $\gamma^*
\to \Pq\bar\Pq$ conversion contribution 
to an integral over the imaginary part of the hadronic vacuum polarization and
thus to the experimentally known quantity
$\Delta\alpha^{(5)}_{\mathrm{had}}(\MZ^2)$.
Therefore no unknown
non-perturbative contribution remains that has to be taken from
experiment.  We also describe practical procedures following
subtraction and phase-space-slicing approaches for isolating and
cancelling the $\gamma^* \to \Pq\bar\Pq$ singularities against the
photon-to-jet conversion function.
The production of Z+jet 
at the LHC is considered as an example, where the photon-to-jet conversion
is part of a correction of the order $\alpha^2/\alphas$ relative to
the leading-order cross section.
\par
\vfill
\noindent 
July 2019 \par
\vskip .5cm 
\null
\setcounter{page}{0}
\clearpage
\def\thefootnote{\arabic{footnote}}
\setcounter{footnote}{0}

\section{Introduction}

The experimental precision for scattering processes at the LHC and
future colliders requires the inclusion of electroweak (EW)
corrections in theoretical predictions. The mixing of EW and QCD
corrections gives rise to additional complications. Since in general 
the leading-order (LO) matrix elements receive contributions of
different orders in the strong and electromagnetic coupling constants,
a complete tower of NLO corrections appears,
as, e.g., discussed for several LHC processes in
\citeres{Dittmaier:2012kx,Frederix:2016ost,Biedermann:2017bss,%
Frederix:2018nkq,Denner:2019tmn}. 
Moreover, the EW
corrections to hadron collider processes involve contributions from
the photon content of the proton, which should be calculated with
photon parton distribution functions (PDFs) based on the LUXqed recipe of
\citeres{Manohar:2016nzj,Manohar:2017eqh}. The photon PDF absorbs
infrared singularities associated with virtual photons coupling to
initial-state particles. The corresponding singularities related to
final-state particles can be treated by using {\it fragmentation
functions}~\cite{Glover:1993xc}.
These are required, in particular, in processes involving photons
and/or jets in the final state, as, e.g., discussed in
\citeres{Denner:2009gj,Denner:2014bna} for 
$\PW+\mathrm{jet}/\gamma$~production at the LHC and for 
jet production in $\Pep\Pem$~annihilation in \citere{Denner:2010ia}.%
\footnote{Alternatively, final-state photons and jets may be isolated
by geometrical cuts that are designed to attribute infrared-singular
contributions to the jets, such as so-called {\it Frixione 
isolation}~\cite{Frixione:1998jh}.}

Beside their direct production, jets can be initiated by
EW mechanisms, in particular via splittings of EW
gauge bosons $V\to f\bar f'$.  For the massive gauge bosons
$V=\PW,\PZ$ those additional jets mostly result from resonant
$\PW/\PZ$~bosons, i.e.\ from process classes that are not directly
related to the ``mother process'' $ab\to C+\mathrm{jet}$ (where $C$ is
any multi-particle final state) and can be treated separately in a
fully perturbative manner.  On the other hand, most mechanisms for
gluonic jet production, $ab\to C+\Pg$, have a direct counterpart in
photon production, $ab\to C+\gamma$, which in turn leads to jet
production via possible splittings $\gamma^*\to q\bar q$ one order
higher in perturbation theory. If the resulting quark- or antiquark-initiated
jets are very close, i.e.\ nearly collinear, they are merged to
one jet by the jet algorithm, so that the resulting event topology
contributes to $ab\to C+\mathrm{jet}$. 
This contribution is infrared
singular in the collinear limit and develops non-perturbative parts, 
since the integration over the virtuality of the
intermediate photon reaches down to the mass scale of the light hadrons
(pions etc.) which is of the order of $\Lambda_{\mathrm{QCD}}$.
By virtue of the KLN theorem~\cite{KLN}
this singularity resulting from real EW corrections to 
$ab\to C+\mathrm{jet}$ could be cancelled by adding the virtual EW 
corrections to $ab\to C+\gamma$ production, similar to the
infrared-safe combination of real and virtual QCD corrections in
the overlap region of one- and two-jet production. 
In experimental analyses, however, the photon production process is 
often separated from the corresponding jet production process.
Hence, the collinear singularity from the low-virtuality limit in the 
$\gamma^*\to q\bar q$ splitting and its accompanying non-perturbative 
contribution do not cancel in cross-section predictions.
Proceeding as in the similar case of identified hadron production,
we absorb the singularity and the non-perturbative contribution
into a ``fragmentation function'' $D_{\gamma\to\jet}$,
which is rather called {\it conversion function} in the following,
because a jet is not an identified hadron.

In the context of EW corrections to LHC processes the fragmentation
functions of quarks and gluons into photons have been used
\cite{Denner:2009gj,Denner:2014bna,Denner:2010ia}. 
These have been introduced in
\citere{Glover:1993xc} and measured by the ALEPH experiment in
photon-plus-jet production at the $\PZ$~pole \cite{Buskulic:1995au}.
Later, the issue of describing the separation of photons and jets in
high-energy collisions via fragmentation functions and their connection 
to EW corrections was briefly outlined in \citere{Frederix:2016ost}
in the context of the calculation of EW NLO corrections to hadronic dijet
production. Here, photon jets are defined as usual using the photon
fragmentation functions $D_{i\to\ga}$. Then, using the
hadron-parton-duality unitarity condition, hadronic jets are defined
as jets that are not photon jets in accordance with the procedure used
in \citere{Denner:2009gj}.

The photon-to-jet conversion function $D_{\gamma\to\jet}$ did not
receive much attention in the literature so far, since its effect,
being of EW origin, is quite small.
Counting the mother process $ab\to C+\Pg$ as ${\cal O}(1)$, the
contribution involving $D_{\gamma\to\jet}$ is suppressed by the
coupling factor $\alpha^2/\alphas$. Nevertheless, this contribution
might compete in size with next-to-next-to-leading-order (NNLO) QCD or
next-to-leading-order (NLO) EW corrections, which involve the relative
coupling factors $\alphas^2$ and $\alpha$, respectively. The simplest
hadronic processes that get contributions from $D_{\gamma\to\jet}$ are
photon-plus-jet and $\PZ$-plus-jet production. More complicated
processes that require such contributions are dijet production, dijet
production in association with a vector boson, and vector-boson
scattering (VBS). For the last process the contribution of
$D_{\gamma\to\jet}$ is actually an
$\mathcal{O}(\alphas)$ correction to the EW VBS process, while it is
still of $\mathcal{O}(\alpha^2/\alphas)$ relative to the LO
contribution to vector-boson-pair + 2\,jet production via strong interactions.
In \citere{Denner:2019tmn}, the NLO QCD and EW corrections to WZ~scattering 
at the LHC, i.e.\ to the EW channel in $\Pp\Pp\to3\Pl\nu+2\,\mathrm{jets}+X$,
were calculated, treating the collinear $\gamma^*\to q\bar q$ 
contribution with the method described in this paper.
  
A lepton collider offers better possibilities to measure the
photon-to-jet conversion function. In photon-plus-jet production 
away from the $\PZ$ resonance peak both
the quark-to-photon fragmentation function and the 
photon-to-jet conversion function contribute at LO.
At LEP this
process has only been investigated on the $\PZ$~pole, where the
contribution of $D_{\gamma\to\jet}$ is strongly suppressed.  Another
possibility is offered by $\PZ$-boson-plus-jet production at lepton
colliders which receives its leading SM contribution exclusively from
the photon-to-jet conversion function and might be suited for a
measurement thereof.
This study could be ideally carried out at some future $\Pep\Pem$~collider
with high luminosity above the Z~resonance.

This paper is organized as follows: In \refse{se:perturbative} we
calculate the contribution of low-virtuality photon transitions to
fermions in perturbation theory. In \refse{se:dispersion} we use a
dispersion relation to express the non-perturbative contribution to
the photon-to-jet transition by the hadronic vacuum polarization. This
result is used in \refse{se:conversion_function} to derive an
approximate result for the photon-to-jet conversion function.
In \refse{se:numerics} we provide an illustrative numerical
application of the photon-to-jet conversion function for
Z+jet production at the LHC.
Our conclusions are presented in \refse{se:conclusion}.

\section{Low-virtuality photon transitions \boldmath{$\gamma^*\to f\bar f$}---perturbative calculation}
\label{se:perturbative}

In perturbative calculations of scattering matrix elements,
contributions appear where a virtual photon splits into a
fermion--antifermion pair. If the virtuality of the photon becomes
small this gives rise to large or singular contributions that require
a dedicated treatment.
Figure~\ref{fig:gammaff} illustrates the leading-order (LO) $\ga^*\to f\bar f$ splitting 
contribution to the cross section for the process $ab\to C+\mathrm{jet}$.
\bfi
\centerline{
 \includegraphics[page=1]{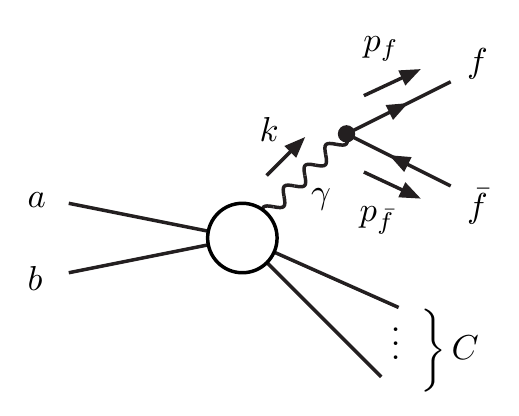}} 
\caption{Generic diagram for the $\ga^*\to f\bar f$ splitting 
contribution to the cross section for the process $ab\to C+\mathrm{jet}$.}
\label{fig:gammaff}
\efi
The definition of the (anti)fermion and photon four-momenta $p_f$,
$p_{\bar f}$, and $k=(p_f+p_{\bar f})$ can also be found there.  In
the phase-space region of low photon virtuality $k^2$, the
contribution to the squared matrix element $|\M_{ab\to Cf\bar
  f}(p_f,p_{\bar f})|^2$ asymptotically factorizes into the squared
matrix element $|\M_{ab\to C\gamma}(\tilde k)|^2$ for a real photon
and a radiator function describing the asymptotic behaviour for
$k^2\to0$ (see, e.g., \citere{Dittmaier:2008md}).  Fully
differentially, spin correlations between the photon and the $f\bar f$
state build up.
But after averaging the splitting process over the
azimuthal angle $\phi_f$ around the collinear axis $\vec k$, the
factorization takes the simple form
\beq
\langle |\M_{ab\to C f\bar f}(p_f,p_{\bar f})|^2 \rangle_{\phi_f}
\;\asymp{k^2\to0}\;
N_{\mathrm{c},f}\,Q_f^2e^2 \, h_{f\bar f}(p_f,p_{\bar f}) \,
|\M_{ab\to C\ga }(\tilde k)|^2,
\label{eq:astff-fact2}
\eeq
where
\beq\label{eq:hff}
h_{f\bar f}(p_f,p_{\bar f}) = \frac{2}{(p_f+p_{\bar f})^2}
\left[ 1-\frac{2}{1-\eps}\left( z(1-z)-\frac{m_f^2}{(p_f+p_{\bar f})^2} \right)\right]
\eeq
and $N_{\mathrm{c},f}$ is the colour multiplicity of
fermion~$f$, i.e.\ $N_{\mathrm{c,lepton}}=1$ and $N_{\mathrm{c,quark}}=3$.
In this asymptotic limit, the virtuality $k^2$ is of the same order as
the square of the light-fermion mass $m_f$, which is assumed to be
much smaller than any relevant scale of the process. For heavy
fermions, the splitting is not enhanced by a singularity since
$(p_f+p_{\bar f})^2>4m_f^2$.  In \refeq{eq:hff}, both the deviation
$\eps=(4-D)/2$ of the number $D$ from the four space--time dimensions
and the non-vanishing fermion mass $m_f$ are kept. Results in dimensional
regularization (DR) for massless fermions or in mass regularization
(MR) in four dimensions can be obtained upon setting $m_f=0$ or
$\eps=0$, respectively.  The energy ratio
\beq
z = \frac{p_f^0}{k^0}
\eeq
controls how the photon momentum $k$ is shared between $f$ and $\bar f$ in the collinear limit,
and the modified photon momentum $\tilde k$ is the on-shell limit ($\tilde k^2=0$)
reached by $k=p_f+p_{\bar f}$ for $k^2\to 0$ in DR or $k^2\to 4m_f^2$ in MR,
where $m_f$ serves just as a regularization parameter.

In \citere{Dittmaier:2008md}, both dipole subtraction functions and the cross-section
contributions in phase-space slicing (defined by a small cut $\Delta\theta$ on the opening angle
between $f$ and $\bar f$) were derived, using the phase-space factorization
described in Sects.~5.1.1 and 5.2.1 of \citere{Catani:2002hc}.
Using the same techniques, it is straightforward to derive the (perturbative)
cross-section contribution of the low-virtuality phase-space region defined by the cut
\beq
4m_f^2 < k^2 < \Delta s
\eeq
on the $f\bar f$~invariant mass $k^2$, which is bounded from below by
the mass threshold for $f\bar f$~production. The cut parameter $\Delta
s$ is smaller than any relevant energy scale 
$Q^2\gg \Delta s$ of the
mother process, but $\Delta s\gg 4m_f^2$ in the case of mass
regularization, where $m_f$ plays merely the role of a regulator.  The
result for the phase-space integral of the squared matrix element is
\beqar
\lefteqn{\int_{k^2<\Delta s}\rd\Phi_{C f\bar f}\,|\M_{ab\to C f\bar f}(\Phi_{C f\bar f})|^2} \\
&=& N_{\mathrm{c},f}\, \frac{Q_f^2\alpha}{2\pi} \,
\int\rd\tilde\Phi_{C\ga}\,
|\M_{ab\to C\gamma }(\tilde k)|^2
\int_0^1 \rd z\,
\Theta_{\cut}\Bigl(p_f=z\tilde k,
p_{\bar f}=(1-z)\tilde k\Bigr)\,
{\cal H}_{f\bar f}(\Delta s,z),
\nn
\label{eq:zintastff1}
\eeqar
which is valid up to terms that are suppressed by the factor $\Delta
s/Q^2\ll1$.  For DR and MR the functions ${\cal H}_{f\bar f}$ are
given by
\beqar
{\cal H}_{f\bar f}^{\mathrm{DR}}(\Delta s,z) &=&
-P_{f\ga}(z) \frac{(4\pi)^\eps}{\Gamma(1-\eps)} \,
\left[\frac{1}{\eps}+\ln\left(\frac{\mu^2}{\Delta s z(1-z)}\right)\right]+2z(1-z),
\\
{\cal H}_{f\bar f}^{\mathrm{MR}}(\Delta s,z) &=&
-P_{f\ga}(z) \ln\left(\frac{m_f^2}{\Delta s z(1-z)}\right)+2z(1-z),
\eeqar
with the $\gamma\to f\bar f$ splitting function
\beq
P_{f\ga}(z) = (1-z)^2 +z^2
\eeq
and $\mu$ denoting the reference mass scale of DR.
The step function $\Theta_{\cut}$ is equal to 1 if an event passes 
all cuts on the momenta $p_f$ and $p_{\bar f}$, and 0 otherwise.
If the complete $z$~range is integrated over, we obtain
\beq
{\int_{k^2<\Delta s}\rd\Phi_{C f\bar f}\,|\M_{ab\to C f\bar f}(\Phi_{C f\bar f})|^2} 
= N_{\mathrm{c},f}\, \frac{Q_f^2\alpha}{2\pi} \,
\int\rd\tilde\Phi_{C\ga}\,
|\M_{ab\to C\gamma }(\tilde k)|^2
{H}_{f\bar f}(\Delta s),
\label{eq:zintastff2}
\eeq
with
\beqar\label{eq:H_DR}
{H}_{f\bar f}^{\mathrm{DR}}(\Delta s) &=&
-\frac{2}{3}\frac{(4\pi)^\eps}{\Gamma(1-\eps)} \,
\left[\frac{1}{\eps}+\ln\left(\frac{\mu^2}{\Delta s}\right)\right]-\frac{10}{9},
\\
{H}_{f\bar f}^{\mathrm{MR}}(\Delta s) &=&
-\frac{2}{3}\ln\left(\frac{m_f^2}{\Delta s}\right)-\frac{10}{9}.
\label{eq:H_MR}
\eeqar
As a technical remark, we note that this collinear singularity
(which does not overlap with a soft singularity)
obeys the simple correspondence 
$(4\pi\mu^2)^\eps/[\eps\Gamma(1-\eps)]\leftrightarrow\ln(m_f^2)$
between the singular terms in DR and MR.

The result of this section can be used to include the low-virtuality
region in a full phase-space integration perturbatively as in any
phase-space slicing approach. Then, the analytical dependence of the
low-virtuality contribution \refeq{eq:zintastff2} on the small cut
parameter $\Delta s$ is cancelled by the implicit dependence of the
remaining phase-space integral on $\Delta s$, which emerges in the
numerical integration, which can be performed for $\eps=0$ and
$m_f=0$.

\section{Low-virtuality photon transitions \boldmath{$\gamma^*\to f\bar f$}---calculation via dispersion relation}
\label{se:dispersion}

The result of the previous section cannot be used directly to evaluate
the low-virtuality contribution to the $ab\to Cf\bar f$ cross section
if $f$ corresponds to quarks. For low virtualities the hadronic
contributions cannot be calculated within perturbation theory as
signalled by the logarithmic quark-mass dependence in MR.
The low-virtuality contribution to the integral $\int\rd\Phi_{Cf\bar
  f}\,|\M_{ab\to Cf\bar f}|^2$ can, however, be evaluated via a
dispersion relation and eventually related to the running
electromagnetic coupling $\alpha(Q^2)$, which is known from low-energy
data on $\Pep\Pem\to f\bar f$, including in particular the case where
the $f\bar f$ states refer to hadrons.

The starting point of this procedure is to rewrite the asymptotic formula for
the squared matrix element in the form
\beq
\langle |\M_{ab\to C f\bar f}(p_f,p_{\bar f})|^2 \rangle_{\phi_f}
\;\asymp{k^2\to0}\;
|\M_{ab\to C\ga }(\tilde k)|^2 \times
\frac{\langle |\M_{\gamma^*\to f\bar f}(k^2)|^2 \rangle}{(k^2)^2},
\label{eq:astff-fact3}
\eeq
where the azimuthal average on the l.h.s.\ can be traded for a photon
spin sum and average in $|\M_{ab\to C\ga }|^2$ and $\langle
|\M_{\gamma^*\to f\bar f}|^2 \rangle$ on the r.h.s., respectively.
Note that the spin-averaged squared matrix element $\langle
|\M_{\gamma^*\to f\bar f}|^2 \rangle$ depends only on the virtuality
$k^2$ and on the splitting variable~$z$, 
but not on the full momenta $p_f$ and $p_{\bar f}$ anymore.
Taking
into account a phase-space factorization over the virtuality $k^2$, we
get
\beq\label{eq:sigma_factorized}
\int_{k^2<\Delta s}\rd\Phi_{C f\bar f}\,
|\M_{ab\to C f\bar f}(p_f,p_{\bar f})|^2 
\;\asymp{k^2\to0}\; \int \rd\widetilde{\Phi}_{C\gamma}\,
|\M_{ab\to C\ga }(\tilde{k})|^2 \times F_f(\Delta s)
\eeq
with
\beq
F_f(\Delta s) =
\int_{k^2<\Delta s}\frac{\rd k^2}{2\pi(k^2)^2}\,
\int\rd \Phi_{f\bar f}\, \langle |\M_{\gamma^*\to f\bar f}(k^2)|^2 \rangle.
\eeq
The phase-space integral over the squared $\gamma^*\to f\bar f$ off-shell
matrix element is related to the imaginary part of the transverse part of the
photon self-energy, $\Sigma^{\gamma\gamma}_{\rT,f}(k^2)$,
via well-known cut equations,
\beq
\int\rd \Phi_{f\bar f}\, \langle |\M_{\gamma^*\to f\bar f}(k^2)|^2 \rangle = 
2\Im\{\Sigma^{\gamma\gamma}_{\rT,f}(k^2)\},
\eeq
where the subscript $f$ in $\Sigma^{\gamma\gamma}_{\rT,f}$ indicates
that only cuts through ``$f$-loops'' (intermediate states involving
the fermion flavour~$f$) are taken into account.
Thus, we get
\beq
F_f(\Delta s) = \frac{1}{\pi} \int_{s'<\Delta s}\rd s'\, 
\frac{\Im\{\Sigma^{\gamma\gamma}_{\rT,f}(s')\}}{s^{\prime2}}.
\eeq
Since $\Sigma^{\gamma\gamma}_{\rT,f}(s)/s$ is an analytic function in
the complex $s$~plane apart from the positive real axis, real and
imaginary parts are related by the dispersion relation
\beq
\frac{\Re\{\Sigma^{\gamma\gamma}_{\rT,f}(s)\}
-s\Sigma^{\prime\gamma\gamma}_{\rT,f}(0)}{s^2} =
\frac{1}{\pi} \Re \int_{4m_\pi^2}^\infty \rd s'\,
\frac{\Im\{\Sigma^{\gamma\gamma}_{\rT,f}(s')\}}{s^{\prime2}(s'-s-\ri 0)},
\label{eq:dispint}
\eeq
where $\Sigma^{\prime\gamma\gamma}_{\rT,f}(0)=
\rd\Sigma^{\gamma\gamma}_{\rT,f}(s)/\rd s|_{s=0}$ 
is a real quantity.
Note that we have used $\Sigma^{\gamma\gamma}_{\rT,f}(0)=0$ because of electromagnetic gauge invariance
and the fact
that $\Im\{\Sigma^{\gamma\gamma}_{\rT,f}(s)\}$
vanishes for $s$ values below the lightest hadronic threshold 
($s<4m_\pi^2$, $m_\pi$ = pion mass)
because of causality.
The running electromagnetic coupling 
\beq
\alpha(s) = \frac{\alpha(0)}{1-\Delta\alpha(s)}, \qquad
\Delta\alpha(s) = \sum_f \Delta\alpha_f(s),
\eeq
comes into play via its
relation to the real part of $\Sigma^{\gamma\gamma}_{\rT,f}$
(see, e.g., \citere{Eidelman:1995ny}),
\beq
\Delta\alpha_f(s) = \Sigma^{\prime\gamma\gamma}_{\rT,f}(0)
-\frac{\Re\{\Sigma^{\gamma\gamma}_{\rT,f}(s)\}}{s}.
\eeq
Note that up to this point all arguments hold to any
order (only the identification of contributions by a flavour $f$ would
deserve clarification beyond NLO). In the following we restrict the
analysis, however, to NLO contributions in the self-energy, which
corresponds to the LO splitting contribution.  The quantity
$\Delta\alpha_{\mathrm{had}}=\sum_q \Delta\alpha_q$ is extracted
\cite{Eidelman:1995ny,Keshavarzi:2018mgv} (see also references
therein) from low-energy data on the ratio
$R=\sigma(\Pep\Pem\to\mbox{hadrons})/\sigma(\Pep\Pem\to\mu^+\mu^-)$
and will be used to evaluate $F_{\mathrm{had}}(\Delta s)=\sum_q
F_q(\Delta s)$.  To this end, we choose $s=\MZ^2\gg\De s$, for which
$\Delta\alpha_{\mathrm{had}}(s)$ is quoted in the literature, and
split the dispersion integral of \refeq{eq:dispint} into a
non-perturbative ($4m_\pi^2<s'<\Delta s$) and a perturbative part
($\Delta s<s'<\infty$),
\beqar
\Delta\alpha_f(\MZ^2) &=&
-\frac{\MZ^2}{\pi} \int_{4m_\pi^2}^{\De s} \rd s'\,
\frac{\Im\{\Sigma^{\gamma\gamma}_{\rT,f}(s')\}}{s^{\prime2}(s'-\MZ^2)}
-\frac{\MZ^2}{\pi} \Re \int_{\De s}^\infty \rd s'\,
\frac{\Im\{\Sigma^{\gamma\gamma}_{\rT,f}(s')\}}{s^{\prime2}(s'-\MZ^2-\ri 0)}
\nn\\
&=&
\frac{1}{\pi} \int_{4m_\pi^2}^{\De s} \rd s'\,
\frac{\Im\{\Sigma^{\gamma\gamma}_{\rT,f}(s')\}}{s^{\prime2}}
- N_{\mathrm{c},f}\, \frac{Q_f^2\alpha}{3\pi} \, \ln\left(\frac{\Delta s}{\MZ^2}\right)
+\dots,
\eeqar
where the non-perturbative part is accurate up to power corrections of 
${\cal O}(M^2_{\mathrm{had}}/\MZ^2)$ with hadron masses $M_{\mathrm{had}}\lsim5\GeV$ 
and the perturbative part up to two-loop corrections.
Thus, we get for $Q^2\gg\Delta s\gg 4m_f^2$ the approximation 
\beq\label{eq:Ffviadalpha}
F_f(\Delta s) = \Delta\alpha_f(\MZ^2) + N_{\mathrm{c},f}\, \frac{Q_f^2\alpha}{3\pi} \, 
\ln\left(\frac{\Delta s}{\MZ^2}\right).
\eeq
Summing over the light quarks (u,d,s,c,b), this yields the hadronic
contribution 
\beq\label{eq:Fhadviadalpha}
F_{\mathrm{had}}(\Delta s) = \Delta\alpha^{(5)}_{\mathrm{had}}(\MZ^2) 
+ \sum_q\frac{Q_q^2\alpha}{\pi} \, 
\ln\left(\frac{\Delta s}{\MZ^2}\right),
\eeq
where the superscript in $\Delta\alpha^{(5)}_{\mathrm{had}}(\MZ^2)$
refers to five active light quark flavours.
This is certainly sufficient to evaluate the ${\cal
  O}(\alpha^2/\alphas)$ corrections induced by the transitions
$\gamma^*\to\mathrm{hadrons}$ at low photon virtualities to any jet
production cross section at the LHC.  A recent fit to
data~\cite{Keshavarzi:2018mgv} gives the result
\beq
\Delta\alpha^{(5)}_{\mathrm{had}}(\MZ^2)= (276.11\pm1.11)\times10^{-4}.
\eeq

To make contact with the fully perturbative calculation of the
previous section, we recall the perturbative NLO expression for
$\Delta\alpha_f(s)$ in MR,
\beq
\Delta\alpha_f(s) = N_{\mathrm{c},f}\,\frac{Q_f^2 \alpha}{3\pi}
\left[\ln\left(\frac{|s|}{m_f^2}\right)-\frac{5}{3}\right],
\eeq
which leads to the perturbative result for $F_f(\Delta s)$,
\beq\label{eq:F_MR}
F_f^{\mathrm{pert,MR}}(\Delta s) = N_{\mathrm{c},f}\, \frac{Q_f^2\alpha}{3\pi} \, 
\left[ \ln\left(\frac{\Delta s}{m_f^2}\right)-\frac{5}{3}\right]
\;=\;
N_{\mathrm{c},f}\, \frac{Q_f^2\alpha}{2\pi} \,
{H}_{f\bar f}^{\mathrm{MR}}(\Delta s),
\eeq
in agreement with the result \refeq{eq:H_MR} of the previous section.
The corresponding result in DR obviously reads
\beq\label{eq:F_DR}
F_f^{\mathrm{pert,DR}}(\Delta s) =
N_{\mathrm{c},f}\, \frac{Q_f^2\alpha}{3\pi} \,
\left[\frac{(4\pi)^\eps}{\Gamma(1-\eps)}\left(-\frac{1}{\eps} + \ln\left(\frac{\Delta s}{\mu^2}\right)\right)-\frac{5}{3}\right]
\;=\;
N_{\mathrm{c},f}\, \frac{Q_f^2\alpha}{2\pi} \,
{H}_{f\bar f}^{\mathrm{DR}}(\Delta s).
\eeq

We conclude this section by a side comment on the cancellation of the
considered singularities as a consequence of the KLN theorem if
photons are considered democratically \cite{Glover:1993xc} as possible
initiators of jets just like any QCD parton. In this case, the cross
section for $ab\to C+\gamma$ becomes part of the $ab\to
C+\mathrm{jet}$ cross section.  Adding the contribution from the
$\gamma^*\to f\bar f$ splitting to the NLO EW cross section for $ab\to
C+\gamma$, 
adds the contribution $\Delta\alpha(Q^2)$ to the relative EW
corrections to this process, where $Q^2$ is some high scale typical
for the process (such as $\MZ^2)$.  Since $\Delta\alpha(Q^2)$ involves
perturbatively ill-defined mass logarithms of the light quarks, the EW
input parameter scheme should be chosen in such a way that those
quark-mass logarithms cancel in the EW correction.  If the
electromagnetic coupling factor $\alpha$ originating from the outgoing
on-shell photon is taken as the fine-structure constant $\alpha(0)$
($\alpha(0)$ scheme), the quark-mass logarithms in the charge
renormalization constant and in the photon wave-function
renormalization constant cancel, so that the additional logarithms in
$\Delta\alpha(Q^2)$ stemming from the photon conversion would remain.
If, however, the respective factor $\alpha$ is effectively taken at
some high scale, as, e.g., in the $\alpha(\MZ^2)$ or $\GF$ schemes \cite{Denner:2000bj,Dittmaier:2001ay,Andersen:2014efa},
the $\Delta\alpha(Q^2)$ contribution from the photon conversion cancels.
In other words, adding the $\gamma^*\to f\bar f$
splitting contribution to the EW correction to the process $ab\to C+\gamma$
effectively replaces the coupling factor 
$\alpha(0)$ for the emitted photon by $\alpha(Q^2)$ 
for some high scale like $Q^2=\MZ^2$.

\section{The photon-to-jet conversion function \boldmath{$D_{\gamma\to\jet}$}}
\label{se:conversion_function}

The common treatment of singular splitting processes associated with
the final state, in which perturbative and non-perturbative
contributions to cross sections arise, makes use of the concept of
fragmentation functions. In the case of the splitting $\ga^*\to q\bar
q$ at low photon virtualities, this means that the NLO cross section
for $ab\to Cq\bar q$ receives a perturbative (pert) contribution, as
calculated above, and a conversion (conv) contribution,
\beq
\sum_q\rd\sigma_{ab\to C q\bar q}(k^2<\Delta s) =
\sum_q\rd\sigma^{\mathrm{pert}}_{ab\to C q\bar q}(k^2<\Delta s) +
\rd\sigma^{\mathrm{conv}}_{ab\to C+\jet},
\label{eq:dsigma}
\eeq
where 
\beqar\label{eq:dsigma_pert}
\rd\sigma^{\mathrm{pert}}_{ab\to C q\bar q}(k^2<\Delta s) &=&
\rd\sigma^{\mathrm{LO}}_{ab\to C\gamma} \, F^{\mathrm{pert}}_q(\Delta s),
\nonumber\\
\rd\sigma^{\mathrm{conv}}_{ab\to C+\jet} &=&
\rd\sigma^{\mathrm{LO}}_{ab\to C\gamma} \, \int_0^1\rd z\, D^{\mathrm{bare}}_{\gamma\to\jet}(z),
\label{eq:dsigma_frag}
\eeqar
and $F^{\mathrm{pert}}_q$ refers to $F^{\mathrm{DR}}_q$
\refeq{eq:F_DR} or $F^{\mathrm{MR}}_q$ \refeq{eq:F_MR} for $f=q$.  Here
$D^{\mathrm{bare}}_{\gamma\to\jet}(z)$ is the ``bare'' $\gamma\to\jet$
conversion function, which depends on the variable $z$ describing
the fraction of the photon momentum $\tilde k$ transferred to one of the
jets \mbox{($p_\jet=z\tilde k$)}.  The bare conversion function
contains singular contributions so that the sum in \refeq{eq:dsigma}
is non-singular. Extracting the singular contribution from
$D^{\mathrm{bare}}_{\gamma\to\jet}(z)$ at some factorization scale
$\mu_{\mathrm{F}}$ requires a ``factorization scheme'', for which we
take the $\MSbar$ scheme following common practice,
\beqar\label{eq:Dbar_DR}
D^{\mathrm{bare,DR}}_{\gamma\to\jet}(z) &=& D_{\gamma\to\jet}(z,\mu_{\mathrm{F}})
+ \sum_q N_{\mathrm{c},q}\, \frac{Q_q^2\alpha}{2\pi} \, \frac{1}{\eps}
\left(\frac{4\pi\mu^2}{\mu_{\mathrm{F}}^2}\right)^\eps \frac{1}{\Gamma(1-\eps)} \, P_{f\ga}(z),
\\
D^{\mathrm{bare,MR}}_{\gamma\to\jet}(z) &=& D_{\gamma\to\jet}(z,\mu_{\mathrm{F}})
+ \sum_q N_{\mathrm{c},q}\, \frac{Q_q^2\alpha}{2\pi} \, 
\ln\left(\frac{m_q^2}{\mu_{\mathrm{F}}^2}\right) \, P_{f\ga}(z).
\label{eq:Dbar_MR}
\eeqar
In DR, it is just the $1/\eps$ pole with the usual prefactors that
is subtracted; in MR we have adjusted the finite contributions
accompanying the singular part ($\propto\alpha\ln m_q$) to define the
same ``renormalized conversion function''
$D_{\gamma\to\jet}(z,\mu_{\mathrm{F}})$ as in DR.  To get a handle
on the non-perturbative contributions to
$D_{\gamma\to\jet}(z,\mu_{\mathrm{F}})$, it would be desirable to
exploit empirical information. This would, however, require an
extremely accurate differential measurement of a jet production cross
section (with low jet invariant mass) and of its corresponding
prompt-photon counterpart, i.e.\ experimental information that is not
available at present.
We can, however, make use of the results of the previous section to at
least get non-perturbative information on
$D_{\gamma\to\jet}(z,\mu_{\mathrm{F}})$ for the case where the full
$z$~range is integrated over.
Comparison of \refeq{eq:sigma_factorized} with
\refeq{eq:dsigma}--\refeq{eq:dsigma_frag} leads to the identification
\beq
F_{\mathrm{had}}(\Delta s) = 
\sum_q F^{\mathrm{pert}}_q(\Delta s) + \int_0^1\rd z\, D^{\mathrm{bare}}_{\gamma\to\jet}(z).
\eeq
Taking the perturbative result for the conversion function  either
in DR \refeq{eq:Dbar_DR} or MR \refeq{eq:Dbar_MR}, and
using \refeq{eq:Fhadviadalpha} and  \refeq{eq:F_MR} or \refeq{eq:F_DR} 
for the integrated renormalized conversion function, we get
\beq
\int_0^1\rd z\, D_{\gamma\to\jet}(z,\mu_{\mathrm{F}}) = \Delta\alpha^{(5)}_{\mathrm{had}}(\MZ^2) 
+ \sum_q N_{\mathrm{c},q}\, \frac{Q_q^2\alpha}{3\pi}  
\left[\ln\left(\frac{\mu_{\mathrm{F}}^2}{\MZ^2}\right)+\frac{5}{3}\right].
\eeq
Note that this $z$-integral of $D_{\gamma\to\jet}$ is sufficient
to evaluate the cross-section contribution $\rd\sigma^{\mathrm{conv}}_{ab\to C+\jet}$
of \refeq{eq:dsigma_frag} with \refeq{eq:Dbar_DR} or \refeq{eq:Dbar_MR}.

The $z$-dependence of $D_{\gamma\to\jet}$ is not provided by the approach
employed in this paper, but would require a model for the hadronization
of the low-virtuality photon into jets. 
At least we can make the following statement on the $z$-dependence
of the conversion function,
\beq
D_{\gamma\to\jet}(z,\mu_{\mathrm{F}}) = \Delta\alpha^{(5)}_{\mathrm{had}}(\MZ^2) \, 
+ \sum_q N_{\mathrm{c},q}\, \frac{Q_q^2\alpha}{2\pi}  
\left[\ln\left(\frac{\mu_{\mathrm{F}}^2}{\MZ^2}\right)+\frac{5}{3}\right]\, P_{f\ga}(z)
+g(z),
\eeq
with $g(z)$ denoting a function that integrates to
$0=\int_0^1\rd z\,g(z)$. 
To reproduce the correct integral over $z$ and thus the correct
cross-section contribution, we can simply set $g(z)\equiv0$, 
\beq
D_{\gamma\to \mathrm{jet}}(z,\mu_{\mathrm{F}}) = \Delta\alpha^{(5)}_{\mathrm{had}}(\MZ^2) 
+ \sum_q  N_{\mathrm{c},q}\,\frac{ Q_q^2\alpha}{2\pi}  
\left[\ln\left(\frac{\mu_{\mathrm{F}}^2}{\MZ^2}\right)+\frac{5}{3}\right]\, P_{f\ga}(z),
\label{eq:Dgjetfinal}
\eeq
in which the non-perturbative $z$-dependence is approximated by a
constant reproducing the correct $z$-integral.

An example for the use of $D_{\gamma\to \mathrm{jet}}$
in some cross-section prediction for the LHC is discussed in
the next section.

\section{\boldmath An example: photon-to-jet conversion 
  function in $\Pp\Pp\to\Pl^+\Pl^-+\jet+X$}
\label{se:numerics}

\begin{figure}
\centerline{\includegraphics[page=2,width=0.21\textwidth]{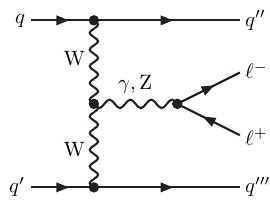}
\quad\includegraphics[page=3,width=0.21\textwidth]{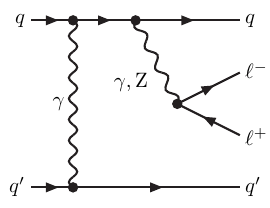}
\quad\includegraphics[page=4,width=0.21\textwidth]{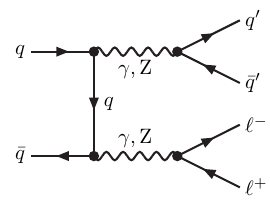}
\quad\includegraphics[page=5,width=0.242\textwidth]{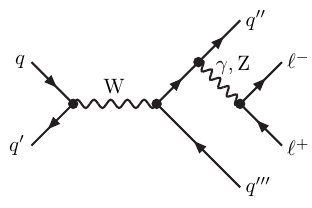}
}
\caption{Some representative Feynman diagrams for
  $\Pq\Pq\to\Pl^+\Pl^-\Pq\Pq$.}
\label{fi:LOdiagrams}
\end{figure}

In this section we focus on the application of the above 
formalism to $\Pp \Pp \to \Pl^+ \Pl^- \Pj+X$.  We consider the
leading-order (LO) cross section at order
$\mathcal{O}\left(\alpha_{\rm s} \alpha^2\right)$.  The contributions
featuring the conversion function are part of the corresponding real
radiation process $\Pp \Pp \to \Pl^+ \Pl^- \Pj \Pj+X$ at order
$\mathcal{O}\left(\alpha^4\right)$ where all QCD partons are quarks.
Some representative Feynman diagrams for this channel are shown in
\reffi{fi:LOdiagrams}. 
While the two quark--quark-induced $t$-channel diagrams on the left of
\reffi{fi:LOdiagrams} dominate the $\mathcal{O}\left(\alpha^4\right)$
contributions, the conversion function only shows up in
quark--antiquark-induced $s$-channel diagrams such as the third
diagram of \reffi{fi:LOdiagrams}.  Moreover, there are channels with
no photon-to-quark conversion at all, as shown in the last diagram of
\reffi{fi:LOdiagrams}.

The numerical study is carried out in the
set-up of \citere{Denner:2011vu}, where the EW corrections of order
$\mathcal{O}\left(\alpha_{\rm s} \alpha^3\right)$ were computed.
We first reproduce the input parameters and the event selection for
completeness and then turn to numerical results.

The simulations are performed for the LHC at $14\TeV$ with the SM
input parameters chosen as
\begin{equation}\arraycolsep 2pt
\begin{array}[b]{lcllcllcl}
\GF & = & 1.16637 \times 10^{-5} \GeV^{-2}, \quad&
\alpha_{\mathrm{s}}(\MZ) &=& 0.1202 , 
&&&
\\
\MW^{\OS} & = & 80.398\GeV, &
\Gamma_\PW^{\OS} & = & 2.141\GeV, \\
\MZ^{\OS} & = & 91.1876\GeV, &
\Gamma_\PZ^{\OS} & = & 2.4952\GeV. &&&\\
\end{array}
\end{equation}
Leptons are considered massless.

Throughout the article, the complex-mass scheme~\cite{Denner:2005fg} is used along with the $\GF$ scheme for $\alpha$.
The on-shell (OS) widths and masses of the W and Z bosons are converted into pole values using \cite{Bardin:1988xt}
\beq
M_V = M_V^{\OS}/
\sqrt{1+(\Ga_V^{\OS}/M_V^{\OS})^2},
\qquad
\Ga_V = \Ga_V^{\OS}/
\sqrt{1+(\Ga_V^{\OS}/M_V^{\OS})^2},
\eeq
leading to the input values
\beqar
\begin{array}[b]{r@{\,}l@{\qquad}r@{\,}l}
\MW &= 80.370\ldots\GeV, & \GW &= 2.1402\ldots\GeV, \\
\MZ &= 91.153\ldots\GeV,& \GZ &= 2.4943\ldots\GeV.
\end{array}
\eeqar
The MSTW2008NLO PDF set~\cite{Martin:2009iq} is used as provided by LHAPDF \cite{Buckley:2014ana}, while
the factorization and renormalization scales are set to the Z-boson mass.

The recombination of QCD partons is done with the $k_{\rT}$-algorithm with $R=0.5$.
The event selection for the numerical analysis is defined as:
\begin{enumerate}
\item Jets are required to have transverse momentum $p_{\rT}$
  larger than $p^{\mathrm{cut}}_{\rT,\mathrm{jet}} = 25\GeV$.
  At least one of them (not necessarily the hardest jet) is required to have
  rapidity $y$ smaller than  $y_{\mathrm{max}} = 2.5$.
\item
The event must have two charged leptons of opposite sign with
transverse momenta $p_{\rT,\ell} > 25\GeV$ and rapidity  $y_\ell < 2.5$.
\item
The dilepton invariant mass is required to fulfil  $M_{\ell\ell} > 50\GeV$.
\item The leptons must be isolated, i.e.\ $R_{\ell\mathrm{jet}} > 0.5$ is
  required for all jets.
\end{enumerate} 

For the simulations, we consider only one lepton family.
In \refta{tab}, we report on the integrated cross section defined in the
fiducial region specified above. The relative corrections of order
$\mathcal{O}\left(\alpha^2/\alphas\right)$ are about half a per cent.
For reference, the EW corrections have been found in
\citere{Denner:2011vu} to amount to a few per cent and the
photon-induced contributions at order
$\mathcal{O}\left(\alpha^3\right)$ to be at the level of $0.1\%$.  The
present findings are in agreement with expectations based on naive
power counting of couplings combined with the fact that the
$\mathcal{O}\left(\alpha^4\right)$ contributions receive some
enhancement owing $t$-channel diagrams in quark--quark channels where
one of the quarks goes into the forward direction (see left two
diagrams in \reffi{fi:LOdiagrams}).
\begin{table}
\begin{center}
\begin{tabular}{cccc}
%
 $\sigma^{\alpha_{\rm s} \alpha^2} [\pb]$ &
 $\sigma^{\alpha^4} [\pb]$  
&  $\delta^{\alpha^4} [\%]$ &  $\delta^{\alpha^4}_\mathrm{conv} [\%]$\\
\hline
$122.414(7)$ & $0.77116(5)$ & $0.63$ & $0.013\%$
%
\end{tabular}
\end{center}
\caption{Cross sections at LO [order $\alpha_{\rm s}\alpha^2$] and 
corrections of order $\alpha^4$ from the real radiation process  
$\Pp \Pp \to \Pl^+ \Pl^-\Pj\Pj+X$ at the $14\TeV$ LHC. The
contribution $\delta^{\alpha^4}_\mathrm{conv}$ of the conversion
function is separately shown for a factorization scale $\mu_{\mathrm{F}}=\MZ$.
The digits in parenthesis indicate the integration error.}
\label{tab}
\end{table}
The contribution of the conversion function is only 
$0.013\%$. Besides the suppression of this contribution by the factor
$\alpha^2/\alphas$ there is an additional suppression due to the fact
that it only features partonic channels with quark--antiquark
initial states (see third 
diagram in \reffi{fi:LOdiagrams}).

\begin{figure}
        \includegraphics[width=0.49\textwidth,page=7]{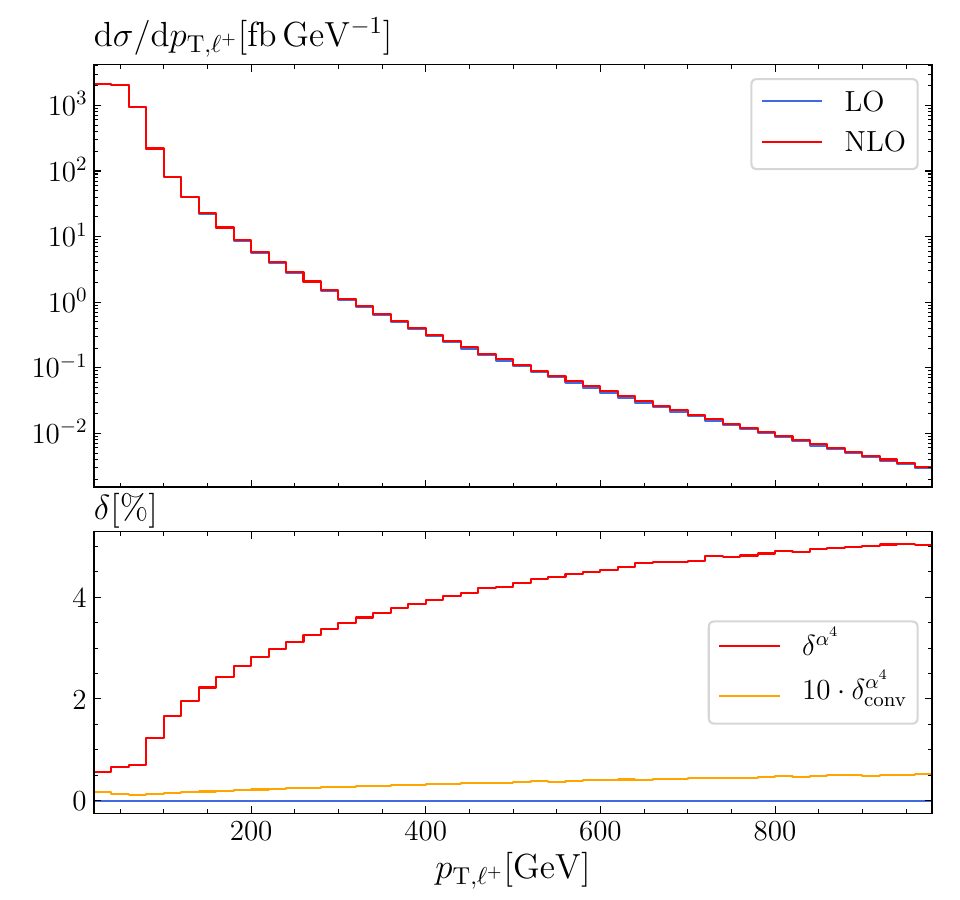}
        \hfill
        \includegraphics[width=0.49\textwidth,page=4]{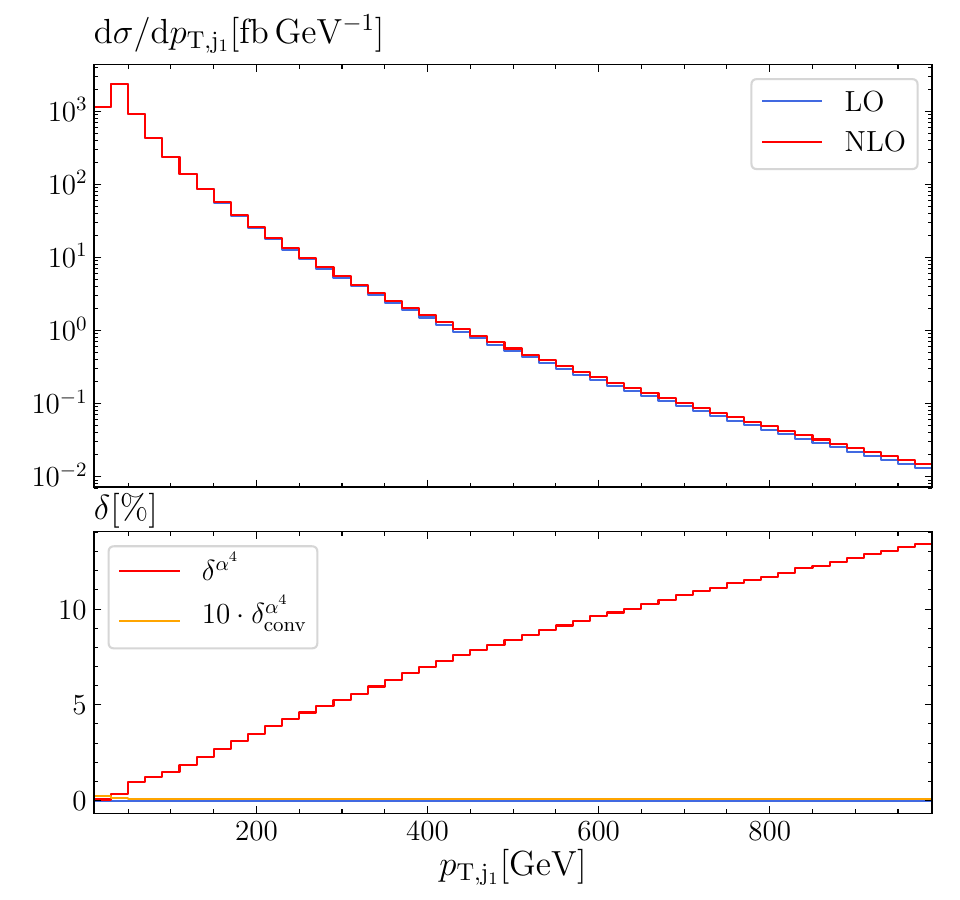}
\caption{\label{dist} Differential distributions for LO [order
  $\mathcal{O}\left(\alpha_{\rm s} \alpha^2\right)$] and corrections
  of order $\mathcal{O}\left(\alpha^4\right)$ from $\Pp \Pp \to \ell^+
  \ell^- \Pj\Pj+X$ at the $14\TeV$ LHC in the  
  transverse momentum of the antilepton (left) and 
  of the hardest jet (right).  The upper panels display the absolute
  predictions, while the lower panels show the relative corrections
of order $\alpha^4$ and its contribution from the photon conversion function.}
\end{figure}

In \reffi{dist}, the differential distributions in the transverse
momentum of the antilepton and the, according to $p_{\rT}$ ordering,
hardest jet are presented.
The corrections  $\delta^{\alpha^4}$ to the transverse momentum of the 
antilepton increase rather smoothly from nearly $0\%$ at the
minimum transverse momentum of $25\GeV$ up to about $5\%$ at $1\TeV$.
For the transverse momentum of the hardest jet, the corrections
increase more strongly and reach more than $10\%$ at $1\TeV$.  This
general trend can be explained by the 
behaviour of the PDFs of the dominant channels.
While the LO contributions [order
$\alpha_{\rm s} \alpha^2$] are dominated by
partonic channels with gluons and quarks in the initial state, the
contributions of the order $\alpha^4$ involve
channels with two valence quarks in the initial state.
The decrease of the gluon PDFs with increasing momentum fraction~$x$
(required by increasing scattering energy)
causes an enhancement of the relative corrections. 
The contribution $\delta^{\alpha^4}_\mathrm{conv}$ of the conversion 
function defined in Eqs.~\refeq{eq:dsigma_frag} and
\refeq{eq:Dgjetfinal} with $\mu_{\mathrm{F}}=\MZ$ is below $0.05\%$
for all considered distributions. 

\section{Conclusion}
\label{se:conclusion}

The calculation of electroweak corrections to processes with jets in
the final state involves contributions of low-virtuality photons
leading to jets in the final state. Such contributions are typically
small but contain infrared singularities, calling for a practical
prescription for their treatment. These singularities can be absorbed
into the photon-to-jet
conversion function, which is similar to a fragmentation function for
identified hadrons.  In this letter, we have used the well-known
hadronic contributions to the vacuum polarization to derive
an approximative expression for the photon-to-jet conversion function.
We have illustrated how this can be used in a practical calculation of
electroweak corrections to Z+jet production at the LHC.

The effect of the photon-to-jet conversion function is typically small
for processes at hadron colliders. Therefore, our recipe is certainly
sufficient for the consistent calculation of electroweak corrections
to processes at the LHC and the next generation of hadron colliders.

A measurement of the photon-to-jet conversion function might be
possible at future high-luminosity lepton colliders in photon-plus-jet
or $\PZ$-boson-plus-jet production above the Z-boson resonance.

\section*{Acknowledgements}

The authors thank the organizers of the Les Houches Workshop 
``Physics at TeV Colliders'', 2019, where this work was
completed, for their kind hospitality and the splendid organization 
of the workshop.
AD acknowledges financial support by
the German Federal Ministry for Education and Research (BMBF) under
contract no.~05H18WWCA1 and the German Research
Foundation (DFG) under reference number DE 623/6-1. 
SD and CS
acknowledge support by the state of Baden-W\"urttemberg through bwHPC
and the DFG through grant no.\ INST 39/963-1 FUGG and grant DI 784/3.
MP is supported by the European Research Council
Consolidator Grant NNLOforLHC2.
CS is supported by the European Research Council under the European
Unions Horizon 2020 research and innovation Programme (grant agreement
no.\ 740006).

\end{document}